\documentclass[11pt]{article}
\pdfoutput=1
\usepackage{jheppub}
\usepackage{subfig}

\newcommand\beq{\begin{equation}}
\newcommand\eeq{\end{equation}}
\newcommand\be{\begin{equation}}
\newcommand\ee{\end{equation}}

\title{Multiband models for field theories with anomalous current dimension}

\preprint{\today}

\author{Andreas Karch}
\affiliation{Department of Physics, University of Washington, Seattle, Wa, 98195-1560, USA}
\emailAdd{akarch@uw.edu}

\abstract{We give simple examples of weakly coupled or free quantum mechanical systems that exhibit scale invariance with an anomalous dimension for a conserved current. In these models scaling as an exact symmetry only emerges in a large $N$ limit, but it remains as an approximate symmetry even at finite $N$.
}

\begin{document}
\maketitle
%\tableofcontents

%%%%%%%%%%%%%%%%%

\section{Introduction}

It has recently become clear that there exists a large class of field theories which have a scaling symmetry under which both the energy density and the charge density have a non-trivial anomalous dimension. This observation has been made in studies of field theories whose dynamics can completely be solved in terms of a holographic dual based on Einstein-Maxwell-Dilaton gravity \cite{Gouteraux:2012yr,Gath:2012pg,Gouteraux:2013oca} or on probe brane constructions \cite{Karch:2014mba,Khveshchenko:2014nka}.  The anomalous dimension of the energy density, as encoded in the hyperscaling violating exponent $\theta$, has long been recognized to be a quite common phenomenon. It occurs, for example, in statistical physics for theories above their critical dimension. A separate anomalous dimension $\Phi$ of the charge density was however unanticipated. There exist even several papers that argue that non-zero $\Phi$ is impossible, for example \cite{wen1992scaling,sachdev1994quantum}. While many of the holographic examples involved bottom-up toy models, some of the theories which produce non-zero $\Phi$ are fairly standard gauge theories in the limit of a large number of colors. For example, the general Dp/Dq system, that is maximally supersymmetric Yang-Mills theories in any spacetime dimension other than 3+1 coupled to matter multiplets in the fundamental representation, preserving half the supersymmetry and potentially localized to lower dimensional planar defects, have been demonstrated to generate non-zero anomalous dimension $\Phi$ for the conserved baryon number current \cite{Karch:2014mba}.

Non-zero $\Phi$ also has potentially interesting applications for the theory of high temperature superconductors. In our earlier work \cite{Hartnoll:2015sea} we demonstrated that transport phenomena in the strange metal phase of the cuprate family can be fitted extremely well, given just a few simple dynamical assumptions, by a critical theory based on dynamical critical exponent $z=4/3$ and $\Phi=-2/3$ together with vanishing hyper-scaling violating exponent $\theta$. The main experiment driving the necessity of non-zero $\Phi$ is the measurement of the temperature dependence of the Hall-Lorenz ratio in \cite{zhang2000determining}. The Lorenz ratio, as well as its Hall version, directly measure the charge of the basic carriers. The fact that this quantity scales in a non-trivial fashion with temperature implies that the charge of the carriers does not act as a constant but has a non-trivial temperature dependence. This is the essence of non-zero exponent $\Phi$. More recent experimental data on the same quantity \cite{matusiak2009enhancement} showed a less clear linear dependence of the Hall-Lorenz ratio as a function of temperature and, in any case, seems to be inconsistent with the findings of \cite{zhang2000determining}. Clearly pinning down this quantity experimentally should be of utmost interest. If one were to interpret the data of \cite{matusiak2009enhancement} as implying a constant Hall-Lorenz ratio, the remaining transport data of the cuprates could be fit with a much more conventional scaling theory \cite{Khveshchenko:2015xea}.

Irrespective of the experimental situation in the cuprates, the question of when non-zero $\Phi$ is consistent or required is clearly of theoretical importance as a basic question in quantum field theory. In our earlier work \cite{Hartnoll:2015sea} we already pointed out potential loopholes in the arguments that seemingly forbid anomalous dimensions for conserved currents. But the strongest evidence for the consistency of non-zero $\Phi$ so far still comes from holography. Note that it is important that the theory is only scale and not conformally invariant, in which case the conformal algebra alone would pin down the dimension of any conserved current to its free value. In the non-relativistic context most scale invariant theories are not conformal.

The fact that up to date the only known examples of non-zero $\Phi$ are based on holography is somewhat disturbing. In this work we are remedying this situation by constructing explicit field theory examples with non-vanishing $\Phi$. The theories we construct will all be ``large $N$", where $N$ is to be thought of as the number of flavors. $N$ can for example count the different bands of a solid. As we will see, scale invariance only emerges as a symmetry in the theories we consider in the large $N$ limit. However, already at moderately large $N$ the properties of the system are very well approximate by the large $N$ scaling answer. The examples are somewhat trivial in the sense that the anomalous dimension for the current appears as a classical phenomenon. It does not arise from divergences in quantum loops, but from summing up an infinite number of contributions from the $N$ flavors in the $N \rightarrow \infty$ limit. This way our theories automatically avoid arguments based on Ward identities that have been put forward to rule out an anomalous dimension for a conserved current.

The organization of this note is as follows. In the next section we present a simple toy model that exhibits how anomalous dimensions arise from large $N$ limits. We give a very simple example of how to obtain a non-vanishing hyperscaling violating exponent out of what is essentially many free systems. In section 3 we give the general construction for non-vanishing $\theta$ based on systems with many flavors (which could be strongly interacting as long as interflavor interactions are suppressed) coupled to general background fields. We give two simple physical examples in section 4. In section 5 we generalize the construction in order to produce a non-vanishing $\Phi$, with a simple physical example for this case in section 6. We discuss finite $N$ corrections in section 7 and conclude with a few comments in section 8.

\section{Hyperscaling violation from a non-relativistic multi-band theory.}
\label{toymodel}

Let us first demonstrate the basic idea of how to get non-trivial scaling exponents from a large number of flavors limit of a multi-band (or multi-flavor) theory in a simple example. For a non-relativistic Fermi gas with dispersion relation $E(p) = p^2/(2m)$ the grand canonical free energy density at zero temperature as a function of chemical potential $\mu$ is given by
\beq
\label{onep}
\omega_0(\mu) = - a \mu^{\frac{d+2}{2}}
\eeq
for positive $\mu$, and it is zero otherwise.
The constant $a$ can easily be determined by filling up the energy levels up to the Fermi energy $E_F=\mu$, but the $\mu$ dependence itself is completely governed by scaling. The system has an underlying scale symmetry with $z=2$ (under which $p$ has dimension 1, $E$ has dimension 2, the mass doesn't scale and the spatial volume has dimension $-d$). Since the free energy density has dimension $d+z$, it has to scale as $\mu^{(d+2)/2}$ as indicated.

A simple generalization of the above model is to include a finite off-set in the dispersion relation,
\beq E(p) = \frac{p^2}{2m} + M \eeq
Clearly all $M$ does is shift the overall energy of all states and the free energy density is given
by\footnote{For $M=0$ the form of $\omega$ in \eqref{onep} implies for energy density $\epsilon$ and particle number density $n$
$$n= \frac{d+2}{2} c \mu^{\frac{d}{2}}, \quad \quad \epsilon = \frac{d}{2} c \mu^{\frac{d+2}{2}}.$$
With finite off-set $M$ the particle number density only sees the difference between chemical potential and off-set, so
$$n= \frac{d+2}{2} c (\mu-M)^{\frac{d}{2}}$$
and similar for the kinetic energy. However, the energy density also receives a direct contribution from the off-set, so that the full energy density is given by
$$ \epsilon = \frac{d}{2} c (\mu-M)^{\frac{d+2}{2}} + n M.$$ Using $\omega=\epsilon-\mu n$, the expression \eqref{withoffset} for the free energy density follow.}
\beq
\label{withoffset}
\omega_0(\mu) = \left \{ \begin{array}{ll} - a (\mu-M)^{\frac{d+2}{2}} & \quad \mbox{for }  \mu > M \\
0 & \quad \mbox{ otherwise}  \end{array} \right . .
\eeq
Note that while $M$ has dimensions of energy and so formally our system is no longer scale invariant, the form of the dispersion relation is still constrained by scaling as long as we account for the fact that $M$ has dimension of energy, dimension $z=2$ that is. The dispersion relation has to take the form
\beq E(M,p) = p^2 f(M/p^2) \eeq
with $f(x)=(1-x)/(2m)$ for the special case of a simple off-set.

Of course for a single band, since $M$ is just some overall shift of all energy levels, we can always set it to zero by a choice of origin. $M$ however becomes meaningful in a multi-band setting, where different bands start at different values of $M$. As an oversimplified example of a multi-band theory let us postulate that we have $N$ flavors of free non-relativistic electrons where the $n$-th flavor has dispersion relation
\beq
E_n(p) = \frac{p^2}{2m} + M_n
\eeq
that is, we take all the flavors to have the same effective mass but different off-sets. We order the flavors by their off-sets, that is $M_{n+1} \geq M_n$. The total free energy density for chemical potential $\mu$ is given by
\beq \label{eps} \omega = -a \sum_{n=1}^{n_{max}} (\mu-M)^{\frac{d+2}{2}} \eeq
where $M_{n_{max}}$ is the largest off-set less than $\mu$.

In the limit of an infinite number of flavors, we can replace the sum over $n$ with an integral:
\beq \omega = a \int_0^{\mu} dM g(M) (\mu-M_n)^{\frac{d+2}{2}} \eeq
where $g(M)$ is the density of flavors, that is $g(M)dM$ counts how many flavors have off-set between $M$ and $M+dM$. This approximation is valid as long as $\mu$ is large compared to the spacing between off-sets.

A very special choice for $g(M)$ is when $g(M)$ is a power law. The simplest case is a constant, that is the off-sets are equally spaced:
\beq g(M) = \frac{1}{m_0} .\eeq
Note that $m_0$ has dimensions of energy; it is exactly the spacing between neighboring off-sets. In this case we can do the integral and obtain
\beq \label{epsans} \omega = \frac{a}{m_0} \int_0^{\mu} (\mu-M)^{\frac{d+2}{2}} = \frac{2 a}{(4+d) m_0} \mu^{\frac{d+4}{z}}. \eeq
That is, we have a free energy density which still has a scale invariance, but this time apparently with hyperscaling violating exponent $\theta=-2$. We will confirm below that this is indeed the correct interpretation. It is important to note that for this special case for $g(m)$ the functional form of the free energy density $\omega$ is still constrained by scale invariance as long as we account for the scale dependence of the single dimensionful parameter $m_0$ which scales like an energy (that is it has dimension $z=2$). Scaling alone guarantees that
\beq
\label{simplescaling}
\omega(\mu,m_0) = \mu^{\frac{d+2}{2}} f(\mu/m_0). \eeq
Now the very fact that we wrote the energy density as an integral over $dM$ with $m_0^{-1}$ only appearing as an overall prefactor immediately tells us that $f(x) \sim x$ and so we can correctly deduce $\omega \sim \mu^{(d+4)/2}$ without even doing the integral. For general power-law density of levels we have (since $g(M) dM$ is dimensionless)
\beq
\label{power}
g(M) = \frac{M^{y-1}}{ m_0^{y} }
\eeq
where $m_0$ is once again a parameter with dimension of energy that characterizes the distribution of levels. Scaling alone still guarantees that the free energy density takes the form \eqref{simplescaling}. Once again, $m_0^{-y}$ appears as an overall prefactor, so we know $f(x) \sim x^y$ and so
 $\omega \sim \mu^{(d+2 + 2 y)/2}$, that is we appear to have hyperscaling violating exponent
\beq \label{thetatoy} \theta = - 2 y. \eeq
This general idea that one can obtain a scale invariant theory by integrating over the mass parameter has also recently been exploited in \cite{pp1,pp2} where it was used to construct an ``unparticle" description of non-Fermi liquids.

\section{General Construction including finite temperature and background fields}
\label{general}

The specific example above demonstrated that in theories with a large number of flavors we can violate naive scaling dimensions and in particular get a free energy density that appears to have a non-zero hyperscaling violating exponent $\theta$. It is very easy to generalize this idea to a generic quantum system or field theory with a large number of flavors, coupled to arbitrary background fields. In particular, we want to turn on a finite chemical potential $\mu$, a finite temperature $T$ and a background vector potential $A_i$ coupled to a conserved particle number current. We assume that each flavor has unit charge under this global particle number symmetry, so that the total current is simply the sum of all the individual flavor currents with equal weight. Each flavor can constitute a strongly coupled system itself, but we assume that the flavors are decoupled from each other so that we can simply get the physical properties of the full system by summing over flavors as above. We assume each flavor is characterized by a parameter $M_n$ with the dimension of energy and the free energy density $\omega$ of each flavor is given by
\beq
\label{individual}
\omega(\mu,T,A_i,M_n) = T^{\frac{d+z}{z}} f(\mu/T,A_i/T^{1/z},M_n/T) \eeq
That is, each flavor is scale invariant with the same dynamical critical exponent $z$ as long as one accounts for the non-trivial scaling of the mass parameter $M_n$. The non-relativistic many-band model of the previous subsection, where $M_n$ was the off-set of the $n$-th band gives a simple example with $z=2$. A theory with a large number of relativistic fermions would be an example with $z=1$. $M_n$ in that case is the mass of the $n$-the flavor. In both cases, the functions $f$ are standard textbook expressions for the free Fermi gas. None of the details of $f$ will be important, other than the fact that $f$ goes to zero faster than $1/M$ at large $M/\mu$. This is expected to be the case as long as the energy of the $n$-th flavor is bigger equal than $M$, so that it's contribution to $f$ is suppressed by a large Boltzmann factor at large $M$. This certainly is true in the case of a free non-relativistic or relativistic gas.

As in our simple warm-up example, in the large number of flavor limit we can get the free energy of the full system by converting the sum over flavors to an integral
\beq
\omega_{tot}(\mu,T,A_i, \ldots) = T^{\frac{d+z}{z}} \, \int_0^\infty dM g(M) f(\mu/T,A_i/T^{1/z},M/T).
\eeq
where the dots stand for all the parameters characterizing the level density $g(M)$.
Returning to the special case that $g(M)$ is a power law as in \eqref{power} characterized by a single quantity $m_0$ carrying dimension of energy we can determine the functional form of $\omega_{tot}$ as above. We know that
\begin{enumerate}
\item $\omega_{tot}$ respects scaling, $\omega_{tot}(\mu,T,A_i,m_0) = T^{\frac{d+z}{z}} F(\mu/T,A_i/T^{1/z},m_0/T)$.
\item The constant $m_0^{-y}$ appears in $\omega_{tot}$ only as an overall prefactor.
\end{enumerate}
This tells us that
\beq
\omega_{tot} = m_0^{-y} \, T^{\frac{d+z}{z} + y} \, \Omega(\mu/T,A_i/T^{1/z}) .
\eeq
This is exactly the statement that the full system has a scale invariance characterized by the same dynamical critical exponent $z$ as the single flavor system but with a hyperscaling violating critical exponent
\beq
\label{theta}
\theta = - y z
\eeq
as in \eqref{thetatoy}. Note that in this expression the background fields $\mu$ and $A_i$ still scale according to their canonical dimensions, that is their anomalous dimension $\Phi=0$.

\section{A simple physical example}
\label{KK}

A theory with an infinite number of flavors may sound fairly exotic, but we can give very simple physical examples of such a many band theory both for the $z=2$ case and the $z=1$ case. The structure of the dispersion relations we require is exactly what one gets from a dimensional reduction. A standard Kaluza-Klein compactification of a free relativistic fermion in $d+1$ spatial dimensions on a circle of radius $R$ gives an infinite tower of flavors in $d$ spatial dimensions of the form we postulated with $M_n=n/R$. Since the masses are all equally spaced this exactly corresponds to the case of a constant density of levels, $y=1$ and so according to \eqref{theta} we have $\theta=-1$ in this case. This is just the statement that when the chemical potential is large compared to the separation of levels, $\mu \gg 1/R$, the system behaves $d+1$ dimensional. From the point of view of the $d$ dimensional theory this appears as a hyperscaling violating exponent $\theta=-1$!

In the non-relativistic $z=2$ system we can accomplish the same effect by confining a $d+1$ dimensional system into a $d$ dimensional quantum well. If we take the confining potential to be an infinite square well of width $L$, we get exactly the many-band theory of our toy example with off-set $M_n \sim n^2/L^2$. Since the distance between levels now grows linearly with $n \sim \sqrt{M_n}$ this corresponds to $g(M) \sim 1/\sqrt{M_n}$ or in other words $y=1/2$. With $z=2$, \eqref{theta} once more yields $\theta=-1$. The hyperscaling violating exponent again simply encodes the higher dimensional character of the theory. Of course the confining potential will in general not take this simple form, but the main point is that we can view any 3d system confined to a quantum well (like the copper oxide layers in the cuprate) as a 2d theory with an infinite number of flavors, so the system is intrinsically ``large $N$" for the purposes of the phenomena discussed here.

\section{Anomalous scaling for conserved currents, electric and magnetic fields}

What we have demonstrated so far is that in systems with many flavors, such as KK reductions or quantum wells, physical quantities can aquire anomalous dimensions from performing the sum over flavors. So far all we accomplished is to obtain a non-trivial hyperscaling violating exponent $\theta$. Non-vanishing $\theta$ has long been appreciated as being an important aspect of critical systems and can be realized without appealing to large $N$ theories, for example in standard critical systems above their critical dimension. A much more puzzling exponent is the anomalous dimension $\Phi$ for conserved currents and consequently for background electric and magnetic field which we recently proposed \cite{Hartnoll:2015sea} to play an important role in the phenomenology of the cuprates based on earlier holographic studies. Holography makes it abundantly clear that $\Phi$ is part of generic critical systems, but no field theory examples with non-vanishing $\Phi$ had been known that did not rely on the holographic duality to determine $\Phi$. We would like to demonstrate that multi-band systems can give us non-vanishing $\Phi$ just as easily as they gave us non-vanishing $\theta$.

From the derivation in section \ref{general} it is clear that the reason we ended up with a vanishing $\Phi$ was that the relative strength with which the various flavors coupled to the background gauge field was equal. All flavors had the same charge. We can generalize our previous construction to flavors with non-equal charge. If we denote the charge of the $n$-the flavor as $e(M_n)$ we see that the free energy of the individual flavor is now given by
\beq
\label{individualtwo}
\omega(\mu,T,A_i,M_n) = T^{\frac{d+z}{z}} f\left( \frac{e(M_n) \mu}{T}, \frac{e(M) A_i}{T^{1/z}},\frac{M_n}{T} \right ) \eeq
in analogy with \eqref{individual}. That is, in the action for the individual flavor $e(M)$, $\mu$ and $A_i$ only appear in the combination $e(M) \mu$ and $e(M) A_i$ and so any dependence on $e(M)$, $A_i$ and $\mu$ can only be in this product form. For the multi-band model we obtain,
\beq
\label{tobeintegrated}
\omega_{tot}(\mu,T,A_i, \ldots) = T^{\frac{d+z}{z}} \, \int_0^\infty dM g(M) \, f\left( \frac{e(M_n) \mu}{T}, \frac{e(M) A_i}{T^{1/z}},\frac{M_n}{T} \right ).
\eeq
For the full theory to still respect any kind of scaling symmetry, we this time need both $g(M)$ and $e(M)$ to be given by power laws
\beq
g(M) = \frac{M^{y-1}}{ m_0^{y} }, \quad \quad e(M) = \frac{M^{\tilde{y}}}{ \tilde{m}_0^{\tilde{y}} }.
\eeq
$m_0$ and $\tilde{m}_0$ are parameters with dimension of energy and the power with which they appear in $g(M)$ and $e(M)$ respectively is determined by the fact that $e(M)$ is dimensionless whereas $g(M)$ has dimension of energy$^{-1}$. Following the logic of the previous sections we can fix the resulting form of the free energy
\beq
\label{integral}
\omega_{tot} = m_0^{-y} T^{\frac{d+z}{z} + y} \, \Omega \left ( \frac{\mu}{T} \left ( \frac{T}{\tilde{m}_0} \right )^{\tilde{y}} ,\frac{A_i}{T^{1/z}}  \left( \frac{T}{\tilde{m}_0}
\right )^{\tilde{y}} \right ) .
\eeq
Concretely, we use that
\begin{enumerate}
\item $m_0$ appears only as an overall prefactor from $g(M)$
\item $\mu$ and $A_i$ show up in the integrand in the combination $\mu/\tilde{m}_0^{\tilde{y}}$ and $A_i/\tilde{m}_0^{\tilde{y}}$ and so they have to appear in this combination in the final answer. $\tilde{m}_0$ only appears in these combinations, so no other powers of $\tilde{m}_0$ occur.
\item The free energy has to respect the scale invariance of the underlying theory with $m_0$ and $\tilde{m}_0$ transforming like energies.
\end{enumerate}
With the standard \cite{Karch:2014mba} assignments $[\mu]=z-\Phi$ and $[A_i]=1-\Phi$ we see that the final answer \eqref{integral} exactly corresponds to the form the free energy should take in a theory with
\beq
\theta = - y z, \quad \quad \Phi=  \tilde{y} z
\eeq

\section{A simple physical realization of non-zero $\Phi$}

As for our theories with non-zero $\theta$, a simple example of a theory which obtains non-vanishing $\Phi$ via the construction outlined in here can be obtained by looking at a Kaluza-Klein example. If we start with a relativistic field that carries charge $q$ already in $d+1$ spatial dimensions, compactification on circle of radius $R$ will give us a tower of particles with mass $n/R$ in $d$ dimensions, every single one of which will carry charge $q$. This is the theory we discussed in section \ref{KK}. The KK-reduction itself however introduces a new $U(1)$ charge in the system. The quantized momentum along the compact direction appears as an extra global $U(1)$ charge in the $d$ dimensional theory. Under this KK $U(1)$ symmetry the particle with mass $n/R$ carries charge $n$. In the language of our construction this corresponds to $y=\tilde{y}=1$, or in other words
\beq \theta=-1, \quad \Phi=1 .\eeq
These dimension assignments can of course easily be understood from the higher dimensional point of view. $\theta$, as before, just signals that an extra dimension opens up. $\Phi=1$ implies that the gauge field has dimension 0 instead of its standard dimension 1. This is to be expected, since the background field $A_{\mu}$ in the $d$ dimensional theory is just the metric component $g_{\mu \phi}$ of the higher dimensional theory, where $\phi$ denotes the compact direction. 0 is indeed the standard dimension assigned to the metric tensor under scaling.

\section{Finite $N$ effects}

\begin{figure}[t]
\subfloat[$N=1$]{\includegraphics[width=2in]{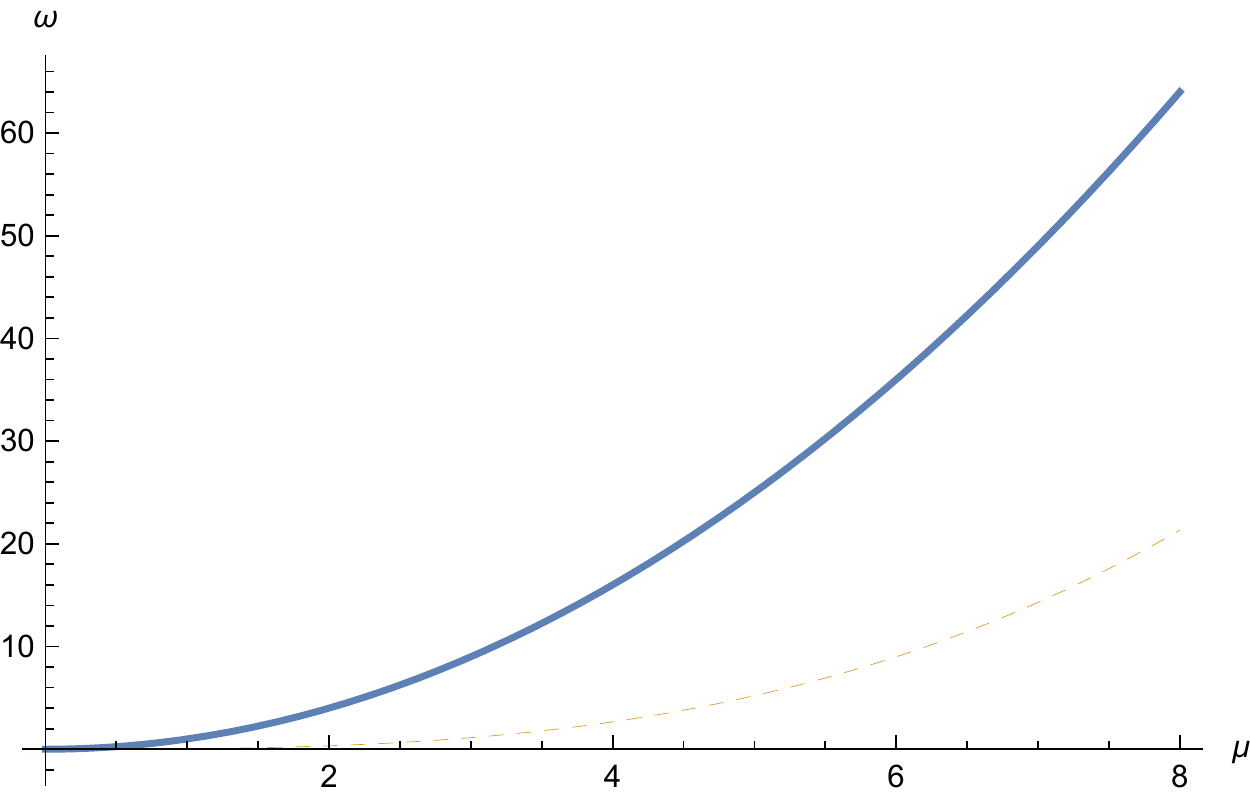}}
\subfloat[$N=10$]{\includegraphics[width=2in]{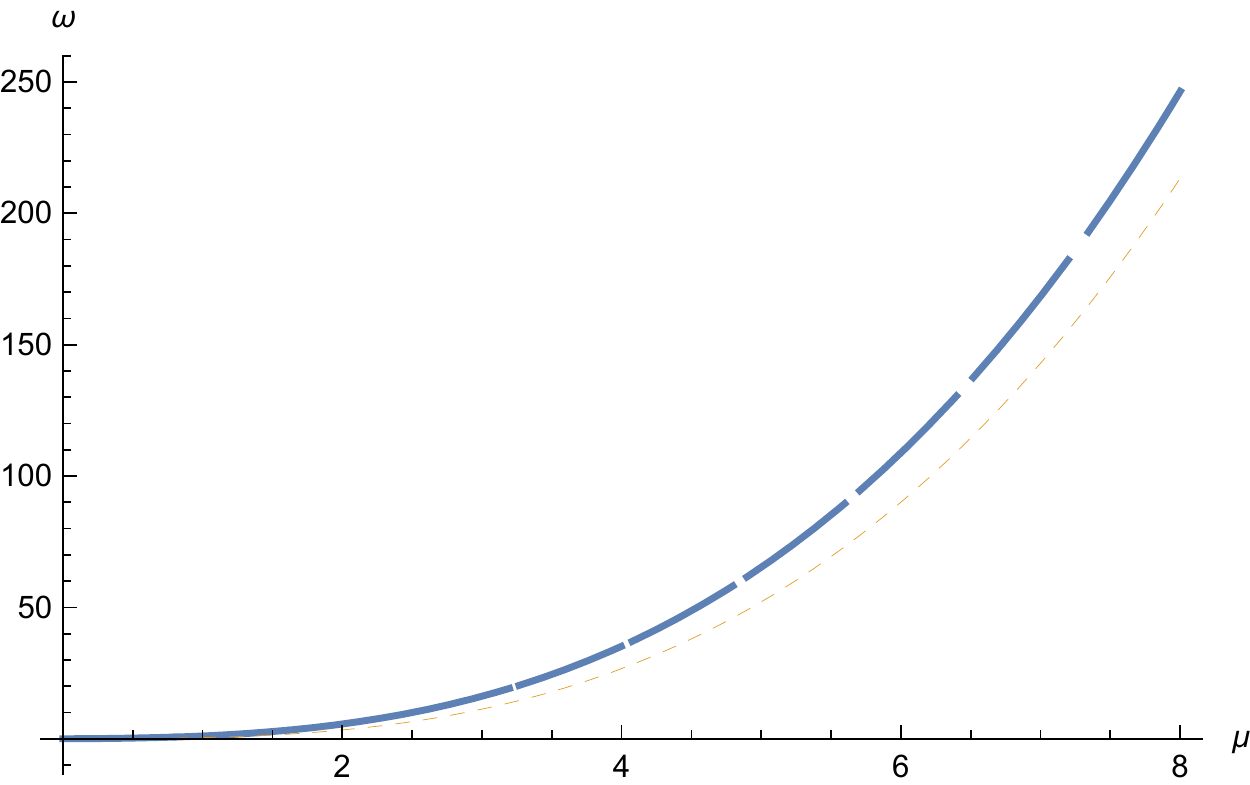}}
\subfloat[$N=20$]{\includegraphics[width=2in]{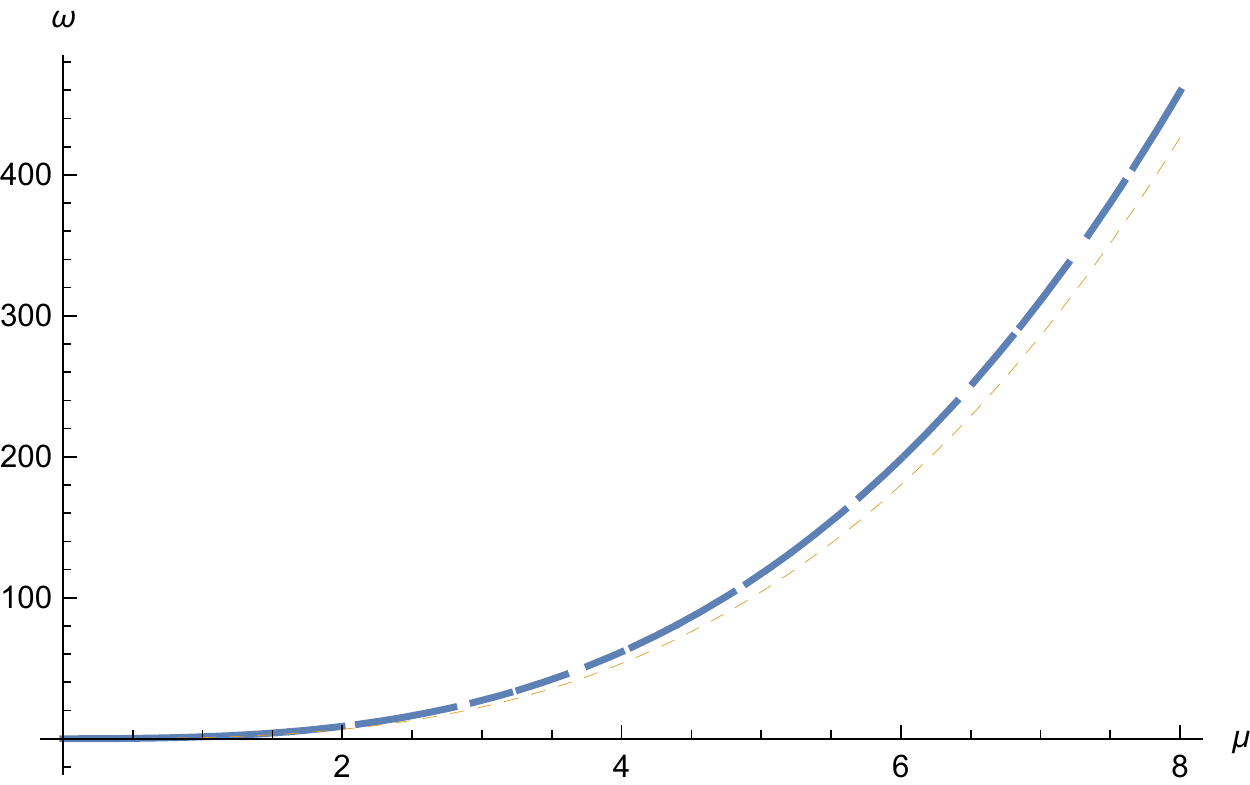}}
\caption{\label{comparison} Free energy for $d=2$ as a function of chemical potential. Depicted is the comparison between the finite sum (solid line) with a) $N=1$, b) $N=10$ and c) $N=20$ levels to the scale invariant answer (dashed line) that emerges at large $N$. The free energy is given by \eqref{eps} with $a=1$. $m_0$ in the continuum answer \eqref{epsans}
 is adjusted to account for the presence of 1, 10, 20 levels in the range of $\mu$ depicted.
}
\end{figure}

Strictly speaking, the scaling symmetry with the non-trivial exponent only emerges in the large $N$, that is in the large number of flavors, limit in the examples we constructed here. However, already at moderately large $N$ is the field theory very well approximated by the large $N$ scaling answer. Scaling with the non-trivial exponents dominates the physics even at finite $N$. To demonstrate this quantitatively, let us return to our simplest toy model of section \ref{toymodel}. In figure \ref{comparison} we compare the finite $N$ answer for 1, 10 and 20 levels to the infinite $N$ scaling answer. While of course for a single level the two disagree wildly one can see that already at the moderately large values of levels the infinite $N$ scaling answer is an extremely good approximation to the full finite $N$ answer.

One important lesson to take away from this study of finite $N$ effects is that the correct notion of $N$ is to count the number of bands within the energy range one wants to study. For scaling to govern the free energy density within a certain range of temperatures or chemical potentials, the number of bands with energy within this range has to be large for the scaling considered in here to be an approximate symmetry.

\section{Comments}
\begin{itemize}
\item These examples easily avoid any theorems based on Ward identities forbidding anomalies for the current. Note that the construction outlined in here would already assign the currents anomalous dimensions {\it classically}. It is the infinite number of flavors/bands that allows currents to pick up an anomalous scaling transformation. The anomalous dimension here is not due to quantum effects
\item We so far neglected interactions between the bands. The flavors within a band can already have arbitrary interactions as long as they do not generate any additional scale. Inter-band interactions will almost certainly renormalize the critical exponents, but since the dimension of the currents was already unconstrained before the interactions are taken into account, any fixed point that emerges in the IR will surely not have to have $\Phi=0$.
\item While it is easy to demonstrate that this construction does give non-vanishing $\Phi$, it is less clear that this is how either holography or cuprates accomplish the feat. Note however that the appearances of many flavors appears natural from the point of view of holography, where the extra dimensions gives infinite towers of states. The fact that cuprates have 2d layers also potentially allows an interpretation in terms of a theory with many flavors, even though in this case the notion that different flavors have different charge still would require quite unusual physics.
\end{itemize}

\section*{Acknowledgements}

We would like to thank Claudio Chamon, Philip Phillips and Larry Yaffe for helpful discussions on related topics. Special thanks to Sean Harnoll for numerous discussions and collaborations on the whole circle of ideas relating to $\Phi$.
This is partially supported by the US Department of Energy under grant number DE-SC0011637.

\bibliographystyle{JHEP}
\bibliography{phi}

\end{document}